\documentclass[conference]{IEEEtran}
\IEEEoverridecommandlockouts
\usepackage{cite}
\usepackage{amsmath,amssymb,amsfonts}
\usepackage{algorithm}
\usepackage{algpseudocode}
\usepackage{graphicx}
\usepackage{braket}
\usepackage{textcomp}
\usepackage{xcolor}
\usepackage[hidelinks]{hyperref}
\def\BibTeX{{\rm B\kern-.05em{\sc i\kern-.025em b}\kern-.08em
    T\kern-.1667em\lower.7ex\hbox{E}\kern-.125emX}}

\begin{document}

\title{Toward Hop-Independent Fidelity in Quantum Data Centers: Resource Requirements for Entanglement Purification\\
}

\author{
\IEEEauthorblockN{
Mohadeseh Azari, Anoosha Fayyaz, Amy Babay, David Tipper, Prashant Krishnamurthy, Kaushik Seshadreesan
}
\IEEEauthorblockA{
\textit{Department of Informatics and Networked Systems} \\
\textit{University of Pittsburgh, Pittsburgh, USA} \\
moa125@pitt.edu}
}

\maketitle

\begin{abstract}
Quantum data-center networks must distribute entanglement between QPUs over paths whose length grows with system scale, but each entanglement-swapping step reduces the fidelity of the raw end-to-end state. Topology, multiplexing, and repeated connection attempts can increase the number of raw end-to-end copies available for a request, yet they do not answer the central resource question: whether those copies are sufficient to remove, via entanglement purification, the fidelity loss caused by multi-hop distribution. We study this question through a topology-independent black-box model of the network. Each elementary link is modeled as a Werner state with parameter $w_0$, so ideal swapping over an $\ell$-link path produces an end-to-end raw copy with Werner parameter $w_0^\ell$; purification succeeds if it outputs at least one state with Werner parameter at least $w_0$ with probability at least $p_{\mathrm{th}}$. We compare recursive BBPSSW purification with higher-order $r$-to-$1$ bilocal-Clifford purification protocols of Jansen \emph{et al.}, using an all-in recursive schedule whose success probability is computed by exact dynamic programming. The resulting resource landscapes show a threshold structure governed by the Werner entanglement condition $w_0^\ell>1/3$ and demonstrate that multi-copy purification substantially improves both feasibility and copy efficiency. Across the evaluated grid, the Jansen family requires fewer copies than BBPSSW at more than $96\%$ of shared feasible points; at $p_{\mathrm{th}}=0.70$, the median copy budget drops from $268$ to $30$. These results provide a quantitative purification-resource benchmark for assessing whether future quantum data-center architectures can practically support hop-independent end-to-end entanglement quality.
\end{abstract}


\section{Introduction}

\begin{figure}[t]
    \centering
    \includegraphics[width=\linewidth]{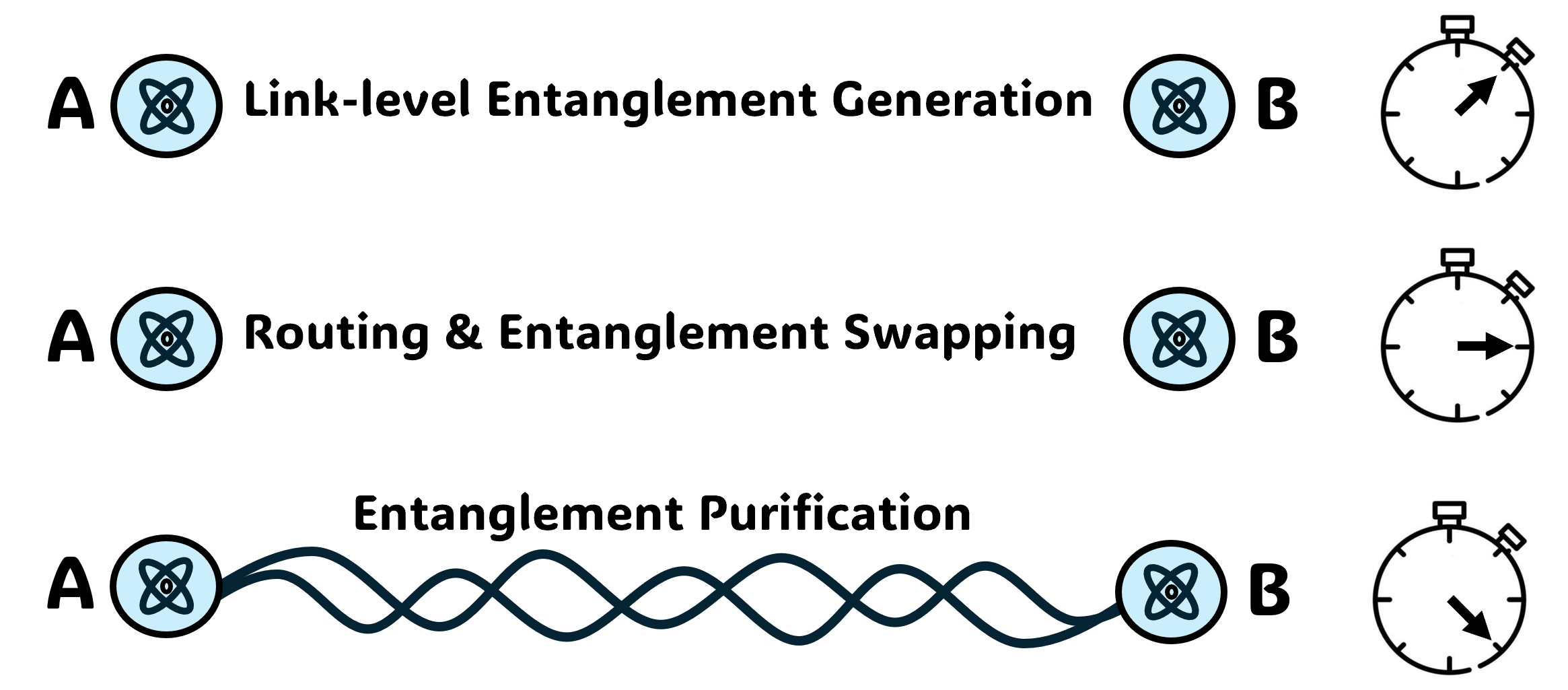}
    \caption{Schematic of the black-box setting studied in this work. First, elementary entanglement is generated across neighboring nodes. Second, routing and entanglement swapping create one or more raw end-to-end entangled pairs between the requested terminal QPUs. Third, entanglement purification should be applied to the available raw end-to-end copies to recover link-level fidelity with a prescribed success probability.}
    \label{fig:three_stages}
\end{figure}

Future Quantum Data Centers (QDC) are expected to rely on architectures in which many quantum processing units (QPUs) communicate through an underlying entanglement distribution network~\cite{Cirac1999Distributed,Caleffi2024Survey,Shapourian2025QDC,Cacciapuoti2025QDC,Pouryousef2026Benchmarking}. This distributed paradigm is motivated by the need to scale beyond the capacity of individual quantum processors by allowing multiple processors to cooperate through shared entanglement. In such systems, scalability depends not only on whether distant QPUs can be connected, but also on whether the delivered entangled states retain sufficient fidelity for distributed quantum tasks. This becomes challenging as the network grows: an end-to-end connection may traverse multiple elementary links, and each swapping operation can reduce the fidelity of the resulting raw end-to-end state \cite{Zukowski1993Swapping,Briegel1998Repeaters,Dur1999Repeaters}. Thus, scalable quantum networking requires a careful understanding of how path length, link-level fidelity, and quantum-state recovery mechanisms interact.

Prior work has shown that network topology, routing, and path diversity can help mitigate rate loss in large quantum networks \cite{Perseguers2013LargeScale,Pirandola2019Capacities,Pant2019Routing,Patil2022DistanceIndependent}. For example, multi-path elementary entanglement generation attempts can increase end-to-end entanglement distribution rates by exploiting network diversity and, under suitable assumptions, can even achieve distance-independent rates. However, rate resilience alone is not sufficient for scalable distributed quantum operation \cite{Briegel1998Repeaters,Kaur2023SquareGrid}. Even if a network can generate many raw end-to-end pairs, the fidelity of each raw pair still degrades with the number of elementary links in the path. Additional copies are useful only if they can be converted into a smaller number of higher-fidelity states.

Entanglement purification provides a natural mechanism for converting multiple copies of imperfect entangled states into fewer copies with higher fidelity. Foundational purification protocols established that noisy entanglement can be probabilistically distilled using local operations and classical communication~\cite{Bennett1996Purification,Deutsch1996QPA}. Purification also plays an important role in quantum repeater architectures, where it is combined with entanglement swapping to counteract fidelity loss over long-distance communication links~\cite{Briegel1998Repeaters,Dur1999Repeaters}. Although the present work focuses instead on multi-hop entanglement distribution inside quantum data-center networks, the same basic tension appears: swapping creates lower-fidelity end-to-end states, and purification requires multiple copies to recover fidelity. The motivation is that the low latency and path diversity of server-centric architectures may allow the topology itself to provide some of the resources required by end-to-end entanglement purification, bringing hop-independent fidelity closer to practical realization.  Recent work on quantum-network purification has examined practical constraints such as noisy operations, finite memories, and scheduling~\cite{Victora2023NetworkPurification,Mobayenjarihani2024Optimistic}, while other studies have developed broader families of multi-copy bi-local Clifford and graph-code-based protocols ~\cite{Jansen2022Enumerating,Goodenough2024GraphCodes}. These lines of work establish the purification mechanisms, yet they do not directly address this paper's central question: before committing to a topology, multiplexing strategy, or scheduling policy, how many raw end-to-end copies does a multi-hop Quantum Data Center network require to achieve fidelity-scalable entanglement distribution? This benchmark can therefore serve as a useful tool for evaluating the fidelity scalability of future server-centric QDC architectures.

We formalize this question through the notion of \emph{hop-independent fidelity} under the Werner-state model. Throughout the paper, we characterize state quality using the Werner parameter, which has a one-to-one affine relationship with fidelity. Let $\ell$ denote the number of elementary links along an end-to-end path, and let $w_0$ denote the Werner parameter of a successfully generated elementary link.Under ideal entanglement swapping, the raw end-to-end Werner parameter over an $\ell$-link path is modeled as
\begin{equation}
    w_{\mathrm{raw}}(\ell) = w_0^\ell .
\end{equation}
We say that a purification procedure achieves hop-independent fidelity for the pair $(w_0,\ell)$ if, using $n_0$ copies of raw end-to-end entangled states of Werner parameter $w_0^\ell$, it produces at least an output state with Werner parameter
\begin{equation}
    w_{\mathrm{e2e}} \geq w_0
\end{equation}
with success probability at least $p_{\mathrm{th}}$. The link-level Werner parameter $w_0$ is therefore the target benchmark. This choice is stricter and more scale-aware than requiring the end-to-end state fidelity to exceed a fixed application-level threshold: the goal is not merely to keep the final state usable, but to remove the quality penalty associated solely with traversing multiple links.

This paper studies hop-independent fidelity in a topology-independent black-box model. Figure~\ref{fig:three_stages} illustrates the operational procedure assumed by this abstraction: elementary entanglement is first generated across neighboring nodes, routing and entanglement swapping then produce raw end-to-end copies for the requested QPU pair, and those copies are finally supplied to the purification stage. In the black-box model, we do not explicitly track how routing is selected, how many paths are used, or how multiplexing, scheduling, and memory constraints interact. Instead, the network core is treated only as a copy-supply mechanism that provides raw end-to-end Werner states over a path of length $\ell$. We then isolate the purification resource requirement itself: for a given link-level Werner parameter, path length, purification protocol, and target success probability, what is the minimum required number of raw end-to-end entangled states to recover link-level Werner parameter? This abstraction gives a necessary resource benchmark for any concrete copy-supply mechanism, including topology-induced path diversity, multiplexed attempts, repeated routing trials, or memory-assisted scheduling.

We compare two classes of purification protocols. First, we use the Bennett \emph{et al.} BBPSSW protocol as the canonical $2$-to-$1$ baseline, since it is one of the foundational entanglement purification protocols and provides a natural reference point \cite{Bennett1996Purification}. Second, we study the higher-order $r$-to-$1$ purification protocols of Jansen \emph{et al.}, which were obtained through bi-local Clifford distillation \cite{Jansen2022Enumerating,Goodenough2024GraphCodes}. Comparing these two families allows us to quantify how strongly the resource cost of hop-independent fidelity depends on the purification strategy.

In this work, we first introduce a black-box formulation of hop-independent fidelity in Werner state model based on link-level Werner-parameter recovery. Second, we compute the minimum number of raw end-to-end copies required to satisfy both the fidelity condition $w_{\mathrm{e2e}}\geq w_0$ and the operational success condition $P_{\mathrm{succ}}\geq p_{\mathrm{th}}$. Third, we evaluate recursive BBPSSW purification and optimized nested $r$-to-$1$ purification under an all-in recursive schedule, using exact dynamic programming to compute the success probability. Finally, we characterize how the required copy budget changes with path length, link-level Werner parameter, purification protocol, and success-probability threshold.

Our results show that the feasibility of hop-independent Werner-parameter recovery is governed primarily by the quality of the raw end-to-end entangled state $(w_0^\ell)$ and by the purification dynamics. The boundary of the feasible region closely tracks the entanglement purification threshold of the input Werner states, while increasing the required success probability increases the number of copies needed within the feasible region. Across the regimes studied, multi-copy $r$-to-$1$ purification substantially reduces the copy requirements relative to recursively applied BBPSSW purification. These results provide a topology-independent resource benchmark for assessing whether future quantum data-center architectures can practically support high-fidelity end-to-end entanglement over multiple hops.

\section{Black-Box Analysis of Hop-Independent Fidelity}
\label{sec:blackbox}

As a first step toward evaluating the resource requirements for hop-independent fidelity, we separate the intrinsic fidelity loss of ideal entanglement swapping, an essential step in any scheme for supplying end-to-end entangled states, from the additional loss introduced by the specific mechanisms used to supply them. Thus, the black-box analysis identifies the minimum resource requirement of a purification scheme. The network is presented only through the resource it provides to the purification layer: a collection of raw end-to-end Werner
states obtained across an $\ell$-link path.

For an elementary-link Werner parameter $w_0$, ideal swapping over $\ell$ links produces raw end-to-end states whose fidelity decreases exponentially with path length. The central question is whether a finite number of such raw copies can be converted into at least one output state whose Werner parameter matches the link-level value $w_0$. We use the minimum required number of raw copies as the resource measure and condition the recovery to succeed
with probability at least $p_{\mathrm{th}}$.

In this section we first define the black-box interface between the network and the purification layer. We then specify the Werner-state assumptions and the purification maps used for both recursive BBPSSW and higher-order $r$-to-$1$ protocols. Third, we describe the dynamic-programming search used to compute the minimum copy budget, and we apply it to determine the minimum resource requirement under an upper bound on the link-level Werner parameter. Finally, we present the feasible region across the space of Werner parameter and end-to-end path length in which hop-independent Werner-parameter recovery is achievable at an acceptable success rate.

\subsection{The Black-Box View}
\label{subsec:blackbox_motivation}

For a requested QPU pair, we condition on the availability of $n_0$ copies of raw end-to-end entangled states generated over paths of similar length $\ell$. Under the ideal-swapping model, these copies are represented as identical Werner states with raw end-to-end parameter $w_0^\ell$. They are therefore equal-quality inputs to the purification stage. The physical and architectural mechanisms that create them---including routing, path diversity, multiplexed link attempts, memory scheduling, and repeated connection trials---are treated in the present black-box analysis as external copy-supply mechanisms rather than as explicitly modeled network processes.

For fixed link-level Werner parameter $w_0$, path length $\ell$, purification protocol, and end-to-end distribution success threshold $p_{\mathrm{th}}$, the black-box problem is to find the smallest copy budget $n_0$ that can recover link-level Werner parameter. The answer specifies the copy-supply requirement that a network implementation would need to meet, whether through topology-induced path diversity, multiplexing, repeated attempts, memory resources, or a combination of these mechanisms. In this sense, the black-box analysis gives a necessary resource benchmark for assessing whether hop-independent fidelity is feasible in a concrete architecture.

The black-box abstraction also separates two distinct failure modes. If no tested purification configuration can raise the raw end-to-end Werner parameter back to $w_0$, then the bottleneck is fidelity recovery: the available input quality is too low for the purification family under consideration. If such a configuration exists but its success probability is small, then the bottleneck is operational: more purification attempts, and hence more raw copies, are required to reach the target success probability. The remaining subsections make these two conditions explicit and compute the
minimum copy budget required to satisfy both.

\subsection{Werner-State Model and Purification Maps}
\label{subsec:blackbox_model}

Each successfully established elementary entanglement link is modeled as a two-qubit
Werner state
\begin{equation}
    \rho_W(w_0)
    =
    w_0 \ket{\Phi^+}\bra{\Phi^+}
    +
    \frac{1-w_0}{4} I_4 ,
    \label{eq:werner_state}
\end{equation}
where $\ket{\Phi^+}$ is a Bell state and $I_4$ is the $4\times4$ identity operator. The parameter $w_0$ is the elementary-link Werner parameter and serves as the target that the purified output state must reach. The fidelity with respect to $\ket{\Phi^+}$ is
\begin{equation}
    F = \bra{\Phi^+}\rho_W(w_0)\ket{\Phi^+}
      = \frac{1+3w_0}{4}.
    \label{eq:werner_to_fidelity}
\end{equation}
A Werner state is entangled when $F>1/2$, or equivalently $w_0>1/3$. The numerical study focuses on the higher-quality operating regime $w_0\in[0.5,1]$.

Under ideal entanglement swapping, the Werner parameter of an end-to-end entanglement distributed over a path of length $\ell$, where $\ell$ denotes the number of elementary links, each having the same link-level Werner parameter $w_0$ along the path, is
\begin{equation}
    w_{\mathrm{raw}}(\ell)=w_0^\ell .
    \label{eq:raw_werner}
\end{equation}
Accordingly, the purification input consists of many of these identical raw end-to-end Werner states, each with parameter $w_{\mathrm{raw}}(\ell)$. The copies are therefore equal-quality inputs corresponding to the same path length in the black-box model. A purification procedure achieves hop-independent fidelity for the pair $(w_0,\ell,n_0)$ if it produces at least one output state satisfying
\begin{equation}
    w_{\mathrm{e2e}} \geq w_0
    \label{eq:werner_recovery_condition}
\end{equation}
with probability at least $p_{\mathrm{th}}$. Here $p_{\mathrm{th}}$ is the operational success threshold, $w_{\mathrm{e2e}}$ the Werner parameter of the purified end-to-end entangled state, and $n_0$ is the raw-copy budget supplied to the purification protocol.

The model is intentionally idealized in order to isolate the resource cost of purification. We assume identical elementary-link Werner parameters, ideal entanglement swapping, ideal local operations inside the purification protocol, and no memory decoherence or storage loss. After each purification step, the output is represented by its Werner-twirled form, so that state quality remains described by a single Werner parameter. These assumptions remove device-level imperfections in order to present only the degradation introduced by multi-hop swapping.

We compare two purification families. The first is the Bennett \emph{et al.} BBPSSW protocol, used as the canonical $2$-to-$1$ recursive baseline~\cite{Bennett1996Purification}. The second is the family of higher-order $r$-to-$1$ protocols of Jansen \emph{et al.}, obtained through bi-local Clifford distillation~\cite{Jansen2022Enumerating}. For the $r$-to-$1$ protocols, we evaluate the output-Werner-parameter and success-probability maps reported in Tables 5 and 6 of Appendix E of Ref.~\cite{Jansen2022Enumerating}. When multiple candidate maps are available for a given block size, the numerical search selects the configuration that minimizes the required copy budget while satisfying both the fidelity and success-probability constraints. The implementation used to generate the numerical results is made available in the accompanying code repository
\cite{HIFCode}.

A single purification round with block size $r$ is represented by two maps,
\begin{equation}
    w^{(r)}_{\mathrm{out}} = f_r(w_{\mathrm{in}}),
    \qquad
    p^{(r)}_{\mathrm{out}} = g_r(w_{\mathrm{in}}),
    \label{eq:generic_purification_map}
\end{equation}
where $w_{\mathrm{in}}$ is the Werner parameter of the input copies, $w^{(r)}_{\mathrm{out}}$ is the Werner parameter of the output copy conditioned on success, and $p^{(r)}_{\mathrm{out}}$ is the success probability of a single $r$-to-$1$ purification. For BBPSSW, $r=2$.

For the BBPSSW baseline, we use the standard Werner-state recurrence.
Given two identical Werner states with input parameter
$w_{\mathrm{in}}$, one successful BBPSSW round produces, after
Werner twirling, an output state with Werner parameter
\begin{equation}
    w^{(2)}_{\mathrm{out}}
    =
    \frac{
        2w_{\mathrm{in}}\left(1+2w_{\mathrm{in}}\right)
    }{
        3\left(1+w_{\mathrm{in}}^2\right)
    },
    \label{eq:bbpssw_werner_update}
\end{equation}
with round success probability
\begin{equation}
    p^{(2)}_{\mathrm{out}}
    =
    \frac{1+w_{\mathrm{in}}^2}{2}.
    \label{eq:bbpssw_success}
\end{equation}

\subsection{Recursive Purification Schedule}
\label{subsec:recursive_trees}

A single purification round probabilistically produces an end-to-end entangled state with higher fidelity than that of its raw input states, but one round may not suffice if the target fidelity for the end-to-end state is that of an elementary link. We therefore allow purification to be applied recursively. The purification level, $k$ ,denotes the number of recursive purification round. At each level, available copies are grouped into disjoint blocks of size $r$, and one $r$-to-$1$ purification attempt is applied to each block. Successful outputs from one level become
the input copies for the next level.

The numerical analysis uses an \emph{all-in} schedule. At each purification level, all currently available copies are used immediately to form as many $r$-copy blocks as possible. Any leftover copies that cannot form a complete block are not used at that level. This differs from partitioning the initial copy budget into independent complete depth-$k$ trees of size $r^k$. The all-in schedule pools all successful outputs after each level before forming
the next set of purification blocks.

This choice is motivated by the recursive structure of purification. The earliest level operates on the lowest-fidelity states, those with Werner parameter $w_0^\ell$, and therefore has the lowest success probability. The all-in schedule uses all available raw copies at this least reliable level, where the greatest number of attempts is needed. As purification proceeds, the number of available input states decreases, but their fidelity—and consequently the success probability of each attempt—increases. The schedule therefore naturally assigns more resources to the less reliable early levels and fewer resources to the more reliable later levels. By contrast, a fixed collection of independent depth-$k$ purification trees partitions the raw copies among complete trees in advance, so only the copies assigned to each tree can contribute to its first-level attempts. The all-in schedule is thus a natural copy-efficient strategy for the black-box resource analysis.

For fixed block size $r$ and recursion depth $k$, the Werner parameter of each purified entangled state evolves deterministically conditioned on successful purification attempt. Starting from the raw end-to-end entangled states, we define
\begin{equation}
    w^{(r,~0)}_{\mathrm{out}} = w_0^\ell ,
\end{equation}
the Werner parameter of the $j^{\mathrm{th}}$-level purification is
\begin{equation}
    w^{(r,~j)}_{\mathrm{out}} = f_r\!\left(w^{(r,~j-1)}_{\mathrm{out}}\right),
    \label{eq:recursive_werner_update}
\end{equation}
and the corresponding single-block success probability at that level is
\begin{equation}
    p^{(r,~j)}_{\mathrm{out}} = g_r\!\left(w^{(r,~j-1)}_{\mathrm{out}}\right),
    \qquad j=1,\ldots,k.
    \label{eq:level_success_probability}
\end{equation}
The final Werner parameter after $k$ levels is
\begin{equation}
    w_{\mathrm{e2e}}(w_0,\ell,r,k)=w^{(r,~k)}_{\mathrm{out}} .
    \label{eq:recursive_output_werner}
\end{equation}
A purification configuration achieves hop-independent fidelity only if
\begin{equation}
    w_{\mathrm{e2e}}(w_0,\ell,r,k)\geq w_0 .
    \label{eq:fidelity_feasible_recursive}
\end{equation}

The success probability is computed by tracking the full probability distribution of the number of copies that survive each purification level. Let $n_j$ be the random variable denoting the number of purified copies after level $j$, with $n_0$ being the initial number of copies available. Suppose that, in the $j^{\mathrm{th}}$ purification round, there are $n_{j-1}$ available copies. The all-in schedule forms $\lfloor n_{j-1}/r\rfloor$ complete purification blocks. Each block succeeds independently with probability $p^{(r,~j)}_{\mathrm{out}}$, because all blocks at level $j$ act on copies with the same Werner parameter $w^{(r,~j-1)}_{\mathrm{out}}$. Therefore, conditioned on $n_{j-1}$, the number of outputs that survive to the next level is binomially distributed:
\begin{equation}
    n_j \mid n_{j-1}
    \sim
    \mathrm{Binomial}\!\left(
        \lfloor n_{j-1}/r\rfloor,
        p^{(r,~j)}_{\mathrm{out}}
    \right).
    \label{eq:level_binomial_transition}
\end{equation}
This equation is the one-level transition rule for the all-in schedule. For example, if $n_{j-1}$ copies are available at level $j$, then only $\lfloor n_{j-1}/r\rfloor$ purification attempts can be made; the possible number of surviving outputs is any integer $n_{j}\in\{0,\ldots,\lfloor n_{j-1}/r\rfloor\}$.

Now lets derive the probability that exactly $n_{j}$ copies are available after level $j$. Using the law of total probability and the binomial transition in Eq.~\eqref{eq:level_binomial_transition}, the probability of having $n_j$ purified entangled copies after level $j$ via recursive $r$-to-$1$ purification is
\begin{equation}
\begin{aligned}
q^{(r,j)}(n_j)
=
\sum_{n_{j-1}}
&q^{(r,j-1)}(n_{j-1})
\binom{\left\lfloor n_{j-1}/r \right\rfloor}{n_j} \\
&\left(p^{(r,j)}_{\mathrm{out}}\right)^{n_j}
\left(1-p^{(r,j)}_{\mathrm{out}}\right)^{
\left\lfloor n_{j-1}/r \right\rfloor-n_j
}.
\end{aligned}
\label{eq:dp_success_recursion}
\end{equation}
Setting $q^{(r,~0)}(n_0)=1$, the summation marginalizes over all possible numbers $n_{j-1}\in\{0,\ldots,\lfloor n_{j-2}/r\rfloor\}$ of copies that could have reached level $j$. For each such $n_{j-1}$, the binomial factor gives the probability that exactly $n_{j}$ of the $\lfloor n_{j-1}/r\rfloor$ purification blocks succeed.
After $k$ purification levels, the protocol succeeds if at least one purified copy remains. Hence the total success probability of the all-in recursive schedule is
\begin{equation}
    p_{\mathrm{e2e}}(n_0;w_0,\ell,r,k)
    =
    \Pr[n_k\geq 1]
    =
    1-q^{(r,~k)}(0).
    \label{eq:all_in_success_probability}
\end{equation}
The condition for hop-independent fidelity is therefore
\begin{equation}
    w_{\mathrm{e2e}}(w_0,\ell,r,k)\geq w_0,
    \quad
    p_{\mathrm{e2e}}(n_0;w_0,\ell,r,k)\geq p_{\mathrm{th}} .
    \label{eq:operational_feasibility}
\end{equation}

This formulation makes the two bottlenecks explicit. The first is a fidelity bottleneck: the recursive map must be capable of raising the raw Werner parameter $w_0^\ell$ back to at least $w_0$. The second is an operational bottleneck: even when the target Werner parameter is reachable, enough raw copies must be supplied so that the all-in schedule produces at least one successful final output with probability at least $p_{\mathrm{th}}$.

As a representative numerical example of the all-in calculation, consider $w_0=0.9327$, $\ell=9$, and $p_{\mathrm{th}}=0.75$. The raw end-to-end input to purification has Werner parameter $w_0^\ell=w^{(r,~0)}_{\mathrm{out}}=0.5343$. For this operating point, the resource-optimized Jansen configuration uses block size $r=4$ and recursion depth $k=2$. Conditioned on success, the first purification level maps the input Werner parameter to $w^{(4,~1)}_{\mathrm{out}}=0.7247$ with single-block success probability $p^{(4,~1)}_{\mathrm{out}}=0.2318$, and the second level maps it to $w^{(4,~2)}_{\mathrm{out}}=0.9327$ with single-block success probability $p^{(4,~2)}_{\mathrm{out}}=0.4188$. Thus, after two purification levels, the output recovers the elementary-link Werner parameter.

The copy budget is determined by the all-in schedule. With $n_0=216$ raw copies, the first level forms $\lfloor 216/4\rfloor=54$ four-copy purification blocks. The number of first-level successes is therefore distributed as $\mathrm{Binomial}(54,p^{(4,~1)}_{\mathrm{out}})$. These successful outputs are then pooled and regrouped into four-copy blocks for the second purification level, whose block success probability is $p^{(4,~2)}_{\mathrm{out}}$. Evaluating the dynamic program in Eq.~\eqref{eq:dp_success_recursion} gives $p_{\mathrm{e2e}}=0.7527$ for $n_0=216$. By contrast, with $n_0=215$ raw copies, only $\lfloor 215/4\rfloor=53$ first-level blocks can be formed, leaving three unused copies at the first level, and the resulting success probability is $p_{\mathrm{e2e}}=0.7452$. Hence, for this $4$-to-$1$u configuration and threshold $p_{\mathrm{th}}=0.75$, the minimum copy budget is $n_0=216$.

\begin{algorithm}[t]
\caption{Minimum-copy search for restoring the end-to-end Werner parameter to at least its link-level value}
\label{alg:min_copy_search}
\begin{algorithmic}[1]
\setlength{\itemsep}{2.5pt}
\Require Link-level Werner parameter $w_0$, path length $\ell$,
end-to-end success threshold $p_{\mathrm{th}}$, allowed block sizes
$\mathcal{R}$, maximum depth $k_{\max}$, maximum exact-DP budget
$n_{0,\max}$
\Ensure Minimum feasible copy budget $n_0^{\min}$ and selected
configuration $(r^\star,k^\star)$

\State $n_0^{\min}\leftarrow\infty$
\State $(r^\star,k^\star)\leftarrow\varnothing$
\For{each allowed block size $r\in\mathcal{R}$}
    \For{$k=1,\ldots,k_{\max}$}
        \State $w_{\mathrm{out}}^{(r,0)}\leftarrow w_0^\ell$
        \For{$j=1,\ldots,k$}
            \State $p_{\mathrm{out}}^{(r,j)}
            \leftarrow
            g_r\!\left(w_{\mathrm{out}}^{(r,j-1)}\right)$
            \State $w_{\mathrm{out}}^{(r,j)}
            \leftarrow
            f_r\!\left(w_{\mathrm{out}}^{(r,j-1)}\right)$
        \EndFor
        \If{$w_{\mathrm{out}}^{(r,k)}<w_0$
        \textbf{or} any $p_{\mathrm{out}}^{(r,j)}=0$}
            \State \textbf{continue}
        \EndIf
        \State Find the smallest $n_0^\star$ via \textsc{BBS} such that
        \[
            \textsc{AllInDP}
            \left(
                n_0^\star,
                \{p_{\mathrm{out}}^{(r,j)}\}_{j=1}^{k},
                r
            \right)
            \geq p_{\mathrm{th}}.
        \]
        \If{no such $n_0^\star$ exists}
            \State \textbf{continue}
        \EndIf
        \If{$n_0^\star<n_0^{\min}$}
            \State $n_0^{\min}\leftarrow n_0^\star$
            \State $(r^\star,k^\star)\leftarrow(r,k)$
        \EndIf
    \EndFor
\EndFor
\If{$n_0^{\min}=\infty$}
    \State Mark the point infeasible within the explored range
\EndIf

\Statex \textbf{Definitions:}
\Statex \hspace{\algorithmicindent}
$\textsc{BBS}$: Bracketed Binary Search
\Statex \hspace{\algorithmicindent}
\textsc{AllInDP}: All-In Dynamic Program

\end{algorithmic}
\end{algorithm}

\subsection{Minimum-Copy Search and Separability Threshold}
\label{subsec:blackbox_search}

\begin{figure*}[t]
    \centering
    \includegraphics[width=\linewidth]{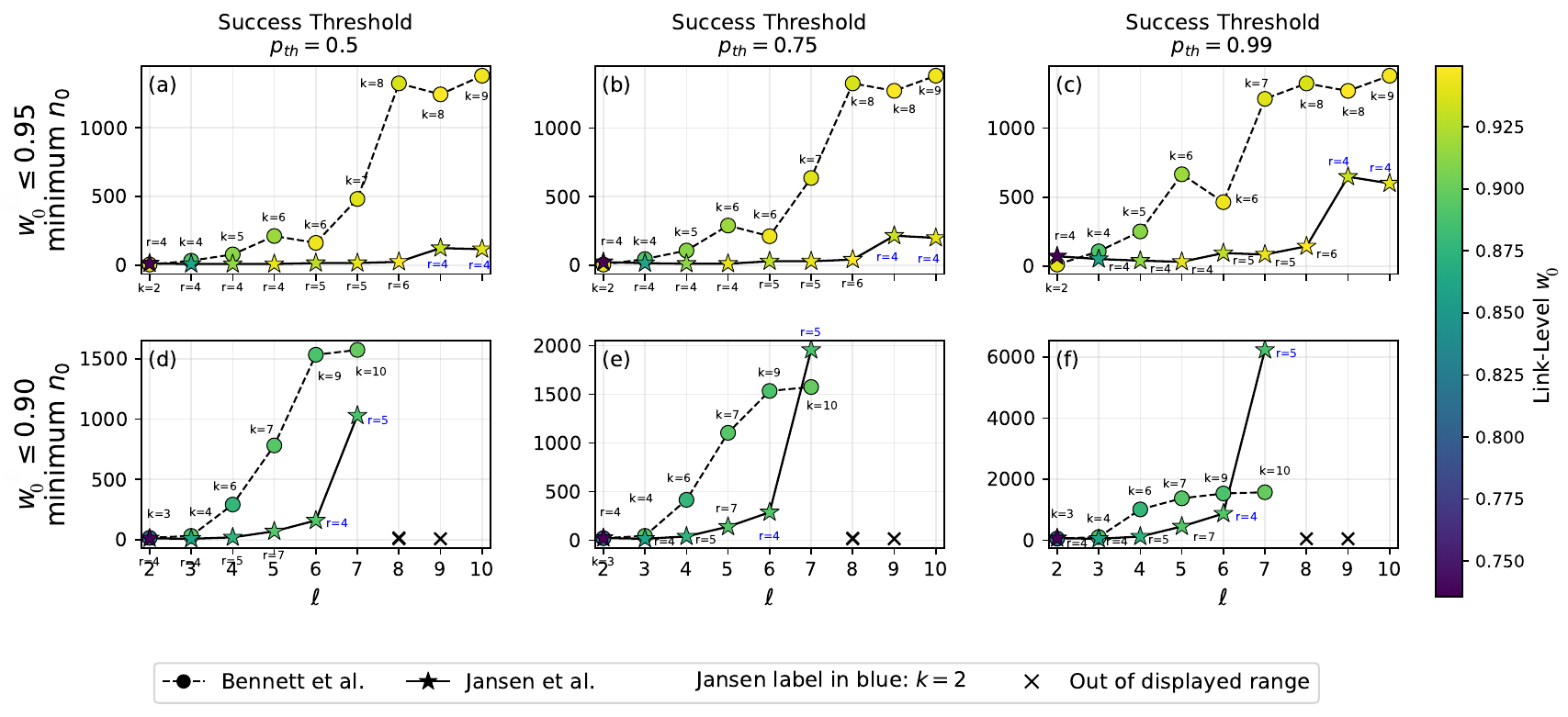}
    \caption{The minimum end-to-end copy budget ($n_0$) required for hop-independent fidelity. Rows correspond to two link-level Werner parameter upper-threshold ($w_{\mathrm{th}}$), and columns correspond to the lower-thresholds on the overall success probability ($p_{\mathrm{th}}$). Circles represent recursive BBPSSW purification, and the labels next to the circles show the selected BBPSSW purification depth ($k$). Stars represent the multi-copy $r$-to-$1$ purification family selected from $r\in\{3,4,5,6,7\}$, and the labels next to the stars show the selected block size $r$. To avoid overcrowding the figure, the selected purification depth is encoded by the color of the $r$-label: black labels indicate $k=1$, while blue labels indicate $k=2$. The y-axis gives the required link-level Werner parameter ($w_0$). Points marked by $\times$ are outside the displayed range.}
    \label{fig:fig3_best_jansen_vs_bbpssw}
\end{figure*}

Given a set of network parameters, each taking a feasible value, the black-box resource problem is to find the smallest raw-copy budget $n_0$ for which at least one of the candidate purification configurations satisfies the hop-independent fidelity condition in Eq.~\eqref{eq:operational_feasibility}. Here, a candidate configuration is specified by two variables: the purification block size $r \in \{2,3,4,5,6,7\}$, which determines the purification protocol, and the recursion depth $k$. For each candidate $(r,k)$, we first evaluate the deterministic Werner-parameter recursion in Eq.~\eqref{eq:recursive_werner_update}. If the resulting output parameter fails to satisfy $w_{\mathrm{e2e}}(w_0,\ell,r,k)\geq w_0$, the configuration is discarded before any copy-budget search is performed. Otherwise, the corresponding level-dependent success probabilities $\{p^{(r,~1)}_{\mathrm{out}},p^{(r,~2)}_{\mathrm{out}},\ldots,p^{(r,~k)}_{\mathrm{out}}\}$ are passed to the all-in dynamic programming recursion of Eq.~\eqref{eq:dp_success_recursion}.

The remaining task is to find the smallest $n_0$ such that $p_{\mathrm{e2e}}(n_0;w_0,\ell,r,k)\geq p_{\mathrm{th}}$. Unlike an independent-tree model, in which each tree independently succeeds with equal per-tree probability, this success probability has no such closed-form expression, because all available copies are used at the first purification level and undergo $k$ levels of probabilistic purification. We therefore evaluate $p_{\mathrm{e2e}}$ directly via dynamic programming for each candidate $(n_0;r,k)$. Since adding more raw copies cannot decrease the probability of producing at least one final purified output, $p_{\mathrm{e2e}}(n_0;w_0,\ell,r,k)$ is non-decreasing in $n_0$. We exploit this monotonicity by bracketing a feasible copy budget and then applying binary search to find the minimum feasible $n_0$.

Algorithm~\ref{alg:min_copy_search} summarizes the numerical procedure. The subroutine \textsc{AllInDP} implements the recursion in Eq.~\eqref{eq:dp_success_recursion} and calculates the success probability from Eq.~\eqref{eq:all_in_success_probability}. A point is marked infeasible if, for every candidate configuration $(r,k)$ in the finite search set, no copy budget $n_0$ within the imposed numerical limit satisfies both the fidelity and success-probability constraints.
We evaluate path lengths from $\ell\in\{2,\ldots,10\}$ and purification block sizes from $r\in\{3,4,5,6,7\}$, with BBPSSW baseline corresponding to $r=2$. The implementation uses exact dynamic programming for the all-in success probability up to the imposed maximum copy budget of $\max{n_{0}} = 50,000,000$. Points requiring copy budgets beyond this range are marked infeasible.

For completeness, we also specify how the two result figures are generated; their trends are interpreted in Sec.~\ref{subsec:blackbox_results}. The compact comparison in Fig.~\ref{fig:fig3_best_jansen_vs_bbpssw} caps the link-level Werner parameter at a threshold $w_{\mathrm{th}}$. For a prescribed elementary-link quality cap $w_{\mathrm{th}}$, path length $\ell$, purification block size $r$, and depth $k$, the link-level target Werner parameter $w_{0}$ must satisfy
\begin{equation}
\begin{gathered}
    w_{\mathrm{out}}^{(r,k)}
    =
    f_r^{\circ k}\!\left((w_{0})^{\ell}\right)
    \geq w_{0},\\
    \text{s.t.}\qquad w_{0}\leq w_{\mathrm{th}}.
\end{gathered}
\label{eq:fixed_target_condition}
\end{equation}
where $f_r^{\circ k}$ denotes applying the purification map $k$ recursive times. For a given $\ell$, we minimize the raw-copy budget $n_0$ over all $(r,k)$ subject to $w_0 < w^{\mathrm{th}}$, selecting the configuration that satisfies the threshold with the fewest copies. Applying this across $\ell$ gives Fig.~\ref{fig:fig3_best_jansen_vs_bbpssw} with color encoding the corresponding $w_0$. In contrast, Fig.~\ref{fig:blackbox_heatmap} evaluates the resource requirement directly over the displayed parameter grid of $(\ell,w_0)$.

The feasibility region is also constrained by a simple analytical boundary. For the Werner state in Eq.~\eqref{eq:werner_state}, the state is entangled only when $w>{1}/{3}$.
Because the raw input to purification has Werner parameter $w_{\mathrm{raw}}=w_0^\ell$, a necessary condition for purification-based recovery is $w_0^\ell>{1}/{3}$,
or equivalently
\begin{equation}
    w_0>3^{-1/\ell}.
    \label{eq:werner_entanglement_boundary}
\end{equation}
Below this curve, the raw end-to-end Werner state is separable, and since LOCC cannot create entanglement, no purification can recover it. Above the curve, recovery is not guaranteed; the purification map must still raise the end-to-end state's Werner parameter back to $w_0$, and the overall success probability must still reach $p_{\mathrm{th}}$. The boundary therefore serves as a necessary reference curve for interpreting the resource landscapes.

\subsection{Resource Scaling Results}
\label{subsec:blackbox_results}

\begin{figure*}
    \centering
    \includegraphics[width=\linewidth]{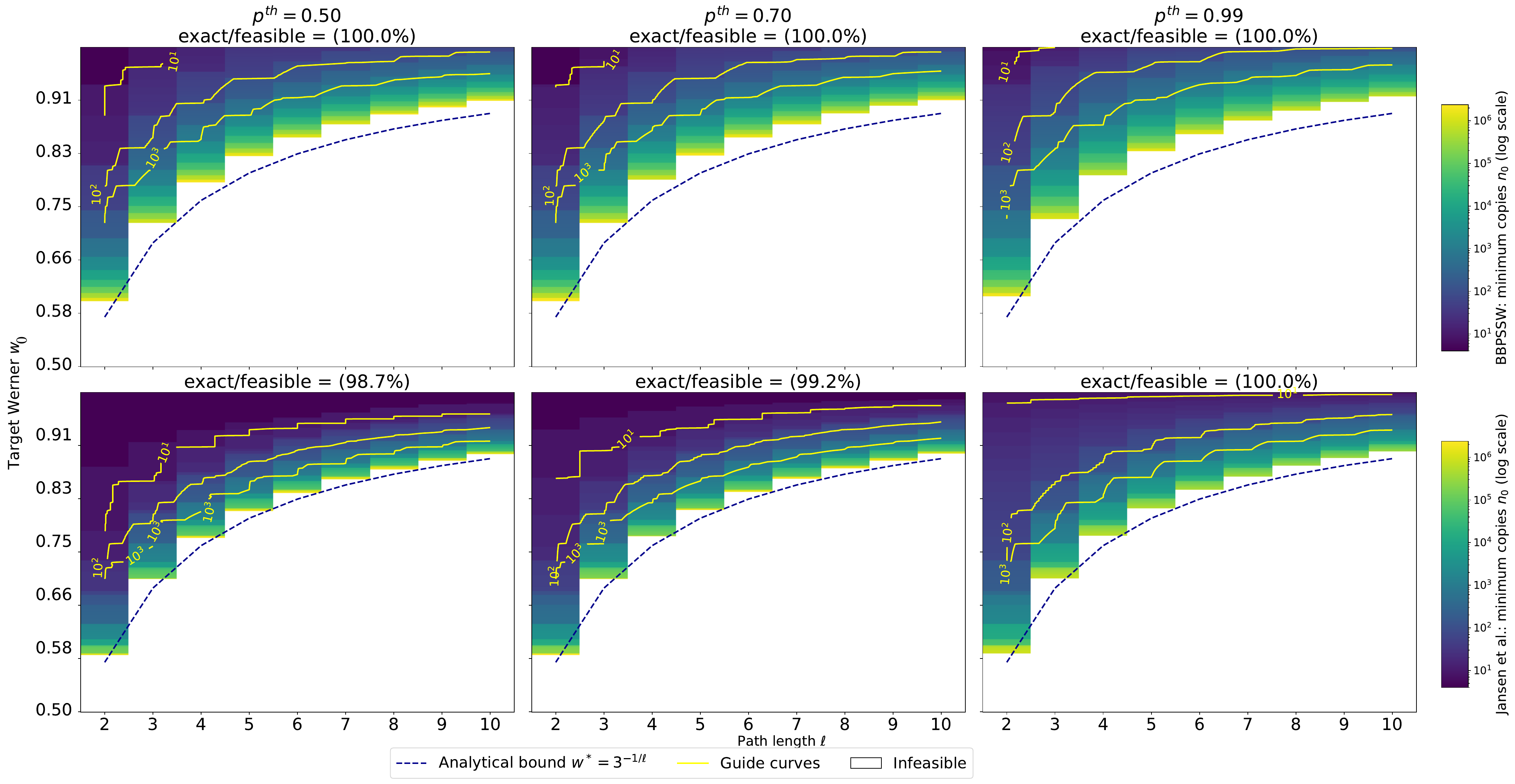}
    \caption{Black-box resource landscape for delivering hop-independent fidelity. Each panel is a heatmap over path length ($\ell$) and elementary-link Werner parameter ($w_0$). The color indicates the minimum end-to-end copy budget ($n_0$) required to produce at least one purified output satisfying $w_{\mathrm{out}}\geq w_0$ with success probability at least $p_{\mathrm{th}}$. The top row uses recursive BBPSSW purification, whereas the bottom row uses the $r$-to-$1$ protocol with $r \in \{3,4,5,6,7\}$ selected to minimize the required resource budget. The dashed curve marks the entanglement separability boundary $w_0=3^{-1/\ell}$. White regions are infeasible within the explored search range, and overlaid contours show constant values of $n_0$.}
    \label{fig:blackbox_heatmap}
\end{figure*}

\begin{figure*}
    \centering
    \includegraphics[width=\linewidth]{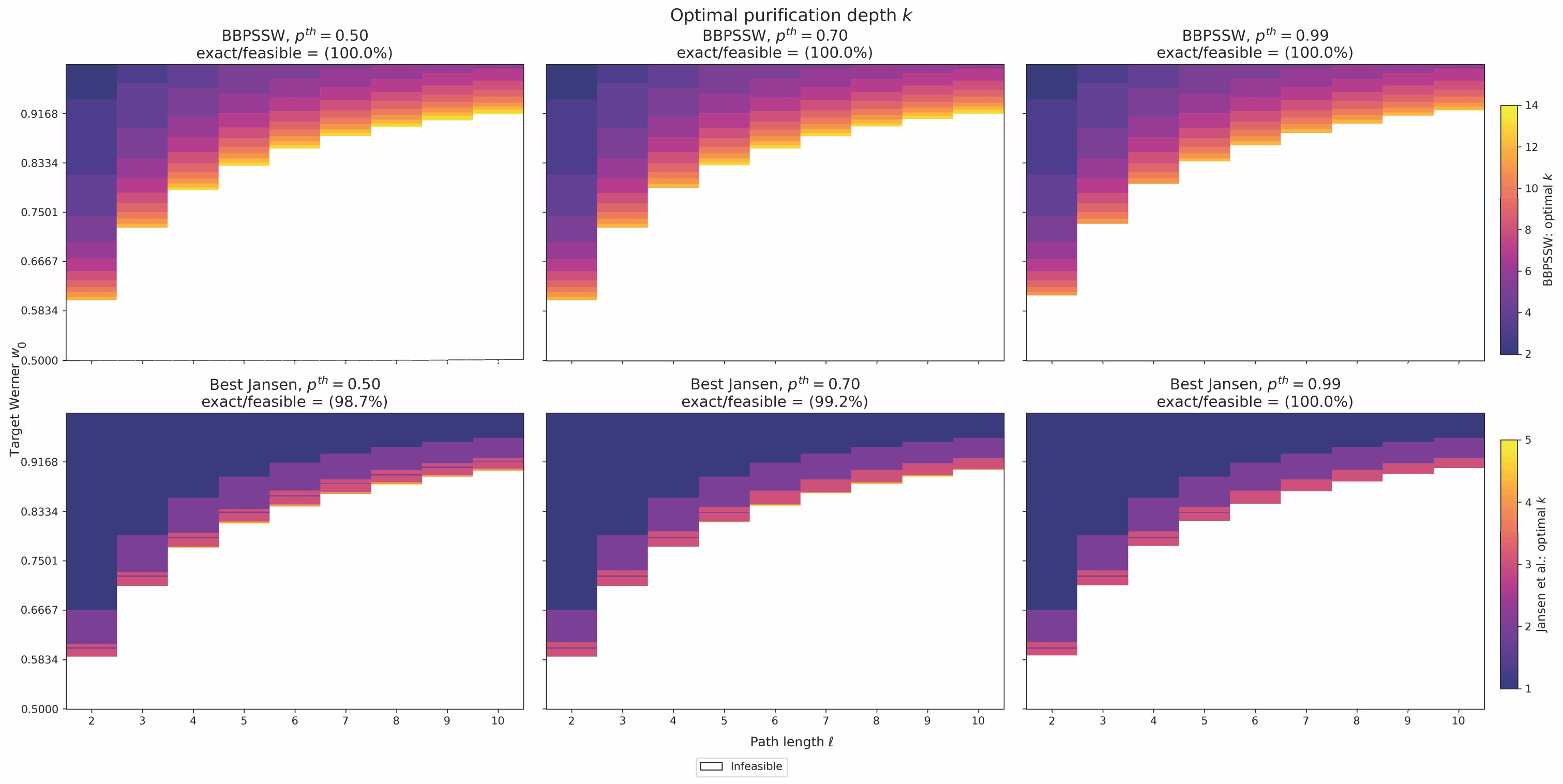}
    \caption{Optimal purification depth, $k$, selected by the minimum-copy optimization underlying Fig.~\ref{fig:blackbox_heatmap}. Each panel is a heatmap over path length ($\ell$) and elementary-link Werner parameter ($w_0$). The color indicates the recursive purification depth selected by the minimum-copy search. The top row corresponds to recursive BBPSSW purification, while the bottom row corresponds to the multi-copy $r$-to-$1$ purification family with $r \in \{3,4,5,6,7\}$ selected to minimize the required resource budget. Columns correspond to the overall success lower-thresholds $p_{\mathrm{th}}\in\{0.50,0.70,0.99\}$. White regions are infeasible within the explored search range.}
    \label{fig:blackbox_k_grid}
\end{figure*}

We now evaluate the resource cost of delivering hop-independent fidelity via end-to-end purification. The results are presented in three complementary figures. Figure~\ref{fig:fig3_best_jansen_vs_bbpssw} compactly compares the baseline $2$-to-$1$ purification with the multi-copy $r$-to-$1$ scheme, showing how the minimum required copy budget varies with path length and overall success threshold at two link-level Werner parameter upper bounds. Figure~\ref{fig:blackbox_heatmap} uses color to represent the minimum required copy budget over the full $(\ell,w_0)$ parameter grid. Figure~\ref{fig:blackbox_k_grid} reports the purification depth ($k$) selected by the same minimum-copy optimization used in Fig.~\ref{fig:blackbox_heatmap}. The top row shows the results of recursive 2-to-1 BBPSSW baseline, while the bottom row gives that of the multi-copy $r$-to-$1$ purification family, where $r$ and $k$ are jointly selected to minimize $n_0$.

The first observation is that the copy budget grows when the input to purification degrades with path length. For fixed elementary-link quality, the raw end-to-end Werner parameter entering purification is $w_0^\ell$, which decreases exponentially as $\ell$ increases. Thus, longer paths push the purification input toward the entanglement separability boundary, and eventually below it if the path is long enough. This directly affects the earliest purification levels: the first level purification act on the raw end-to-end state and has the lowest success probability $p^{(r,~1)}_{\mathrm{out}}=g_r(w_0^\ell)$ in the recursive schedule. As a result, longer paths typically cause greater attrition at the beginning of the all-in purification process. Consequently, hop-independent recovery over longer paths requires deeper recursion, larger purification block sizes, or both, and thus demands more raw end-to-end resources.

Figure~\ref{fig:fig3_best_jansen_vs_bbpssw} shows that this growth is not strictly monotone. The minimum-copy search is an optimization over both the purification block size ($r$) and the recursion depth ($k$). As $\ell$ changes, the optimal purification setting satisfying the hop-independent fidelity constraints of Eq.~\ref{eq:operational_feasibility}, can switch from one configuration to another. These transitions are indicated directly in the figure. For BBPSSW, the labels next to the circle markers show the selected recursion depth $k$. For the multi-copy purification, the labels next to the star markers show the selected block size $r$; because the optimized depths were only $k=1$ or $k=2$, the depth is encoded by the color of the $r$-label, with black indicating $k=1$ and blue indicating $k=2$. When the link-level Werner parameters are higher (the top row), multi-copy purification typically uses a single-level $r=4$ setting for shorter and moderate path lengths, then switches to larger block sizes or to $k=2$ recursion as the path length increases. When link-level Werner parameter is moderate (the bottom row), deeper recursion becomes necessary earlier because the raw end-to-end Werner parameter is closer to the separability threshold.

The comparison with BBPSSW highlights the value of multi-copy purification. BBPSSW can be competitive for very short paths, where the raw end-to-end Werner parameter is still high and the overhead of a larger purification block is not yet justified. However, this advantage disappears quickly as the path length grows. Once the fidelity of the raw end-to-end state has degraded enough that repeated $2$-to-$1$ purification becomes costly, multi-copy purification offers a more efficient route to boosting the end-to-end Werner parameter, using larger purification blocks in a single round or a few recursive rounds.

This behavior is evident in Fig.~\ref{fig:fig3_best_jansen_vs_bbpssw}. The optimized multi-copy configuration reduces the copy budget by large factors for moderate and longer paths; at $w_{th}=0.95$ and $\ell=8$, for instance, it requires $42$ copies versus $1324$ for BBPSSW. The savings, however, shrink at shorter paths, where the raw end-to-end state is still high-fidelity and BBPSSW smaller blocks consume fewer copies per round than the larger multi-copy blocks, and at very high success thresholds, where the all-in schedule must allocate many additional raw copies to meet the required success probability. Overall, the trend is that multi-copy purification is most valuable precisely in the multi-hop regime where hop-independent recovery is hardest.

Figure~\ref{fig:blackbox_heatmap} extends the sweep over path length and elementary-link Werner parameter, with the dashed curve marking the separability boundary. Below this boundary, the raw end-to-end Werner state is separable; above it, the state is entangled, but purification must still satisfy the hop-independent fidelity condition in Eq.~\eqref{eq:operational_feasibility}. The boundary therefore shifts upward with path length, as stronger elementary-link entanglement is required to preserve end-to-end entanglement. The transition from the colored to the white region marks the point at which the required copy budget exceeds the search cap. At $p_{\mathrm{th}}=0.5$ ($0.99$), its average gap from the analytical boundary is $0.010$ ($0.014$) for multi-copy purification, compared with $0.025$ ($0.032$) for BBPSSW. Because these boundaries occur at very large copy budgets, their relative proximity to the analytical limit indicates greater numerical efficiency rather than practical achievability.

The full sweep reveals the principal advantage of multi-copy purification: its substantially lower copy requirement across the operating region. Over the entire $(\ell,w_0)$ heatmap, multi-copy purification is feasible across a larger portion of the parameter space than BBPSSW for every tested end-to-end success threshold. At $p_{\mathrm{th}}=0.5$, BBPSSW is feasible at $1444$ grid points, while the optimized multi-copy configuration is feasible at $1565$; at $p_{\mathrm{th}}=0.99$, the corresponding counts are $1386$ and $1534$, respectively. Moreover, whenever both protocol families are feasible, multi-copy purification requires a lower copy budget at more than $96\%$ of the shared feasible grid points for every tested success threshold. This advantage is particularly pronounced in the low-copy regime: at $p_{\mathrm{th}}=0.5$, multi-copy purification requires at most $10$ initial raw copies at $656$ grid points, compared with only $125$ for BBPSSW. Thus, the resource advantage of multi-copy purification is not restricted to a small corner of the parameter space but is broadly present across the operating region, including configurations with comparatively modest copy requirements.

Unlike path length, the end-to-end success threshold primarily affects the number of required copies rather than the selected purification configuration. Increasing $p_{\mathrm{th}}$ from $0.5$ to $0.99$ raises the copy budget throughout the feasible region, while the feasibility boundary and the selected recursion-depth patterns in Fig.~\ref{fig:blackbox_k_grid} remain largely unchanged. At fixed $(\ell,w_0)$, changing $p_{\mathrm{th}}$ does not alter the raw end-to-end Werner parameter $w_0^\ell$ or the fidelity evolution under purification. Consequently, once a minimum-copy purification configuration has been selected, a stricter success requirement is generally met by supplying more raw copies to the same all-in schedule. By contrast, increasing $\ell$ changes $w_0^\ell$ and can therefore trigger a switch in the selected purification configuration, producing the non-monotonic copy-budget behavior observed in Fig.~\ref{fig:fig3_best_jansen_vs_bbpssw}. 

The median copy budgets across the feasible heatmap points summarize this overall increase. At $p_{\mathrm{th}}=0.5$, the median is $220$ for BBPSSW but only $20$ for the multi-copy purification family; at $p_{\mathrm{th}}=0.70$, the corresponding medians are $268$ and $30$; and at $p_{\mathrm{th}}=0.99$, they increase to $607$ and $100$, respectively. Thus, $p_{\mathrm{th}}$ primarily controls the resources needed to execute the selected purification schedule with the required reliability, whereas $(\ell,w_0)$ primarily determines the structure of that schedule, allowing the relative resource advantage of multi-copy purification to remain substantial as the success requirement increases.

\textcolor{purple}{Here}
Figure~\ref{fig:blackbox_k_grid} provides a complementary view of the multi-copy advantage by showing the purification depth selected by the minimum-copy optimizer. BBPSSW generally requires substantially deeper recursion: across the feasible heatmap points, its median selected depth is $k=6$ for all three success thresholds, with depths reaching $k=14$ for $p_{\mathrm{th}}=0.50$ and $p_{\mathrm{th}}=0.70$. By contrast, the multi-copy family has a median depth of $k=1$ across all three thresholds, with only a small fraction of the grid requiring depths greater than $k=2$. Moreover, at every shared feasible grid point, the selected multi-copy configuration is shallower than BBPSSW for each tested end-to-end success threshold. Thus, the multi-copy advantage is reflected not only in a lower initial copy budget $n_0$, but also in shallower purification schedules for reaching the target Werner parameter.

The selected purification block sizes reveal how the optimizer uses the available multi-copy protocols. Across the heatmap grid, $r=4$ is the dominant choice, while $r=5$ is the next most common; $r=6$ and $r=7$ appear mainly in larger path length regimes. This pattern suggests that most of the gain over BBPSSW comes from moving beyond pairwise purification to moderate-size multi-copy purification, rather than from always using the largest available block size. Larger $r$ can help near difficult operating points, but it also increases the number of raw copies consumed by each purification attempt. The optimal protocol therefore balances stronger one-round purification against the all-in schedule's need to maintain enough surviving copies through the recursive levels.

Taken together, the results support three conclusions. First, hop-independent fidelity can be achieved through end-to-end purification only if the raw end-to-end Werner parameter exceeds the input-quality threshold of the selected purification configuration. Second, the end-to-end success threshold mainly controls the required copy budget after fidelity recovery becomes possible, rather than redefining the optimized purification setting. Third, higher-order multi-copy purification substantially improves both feasibility and copy efficiency relative to recursive BBPSSW, and Fig.~\ref{fig:blackbox_k_grid} shows that this improvement is accompanied by much shallower optimal purification schedules.

\section{Discussion}
\label{sec:discussion}

The black-box analysis in this paper deliberately separates the purification-side resource requirement from the network-side mechanisms that supply raw end-to-end copies. This separation makes the minimum copy budget transparent, but it also leaves several important system-level questions open. The most immediate next step is to open the black box and ask whether concrete quantum data-center network architectures can supply the required end-to-end raw copies efficiently. Different architectures may do so in different ways: path diversity may provide parallel end-to-end copies, multiplexed elementary links may increase the number of successful raw states per time window, repeated routing attempts may accumulate copies over time, and quantum memories may store successful states until enough inputs are available for purification. A topology-aware extension of this work would therefore compare candidate data center topologies by the copy budgets they can support as a function of path length, elementary-link success probability, memory capacity, and scheduling policy.

A natural topology-aware continuation is to instantiate the copy-supply side of the model in concrete Quantum Data Center (QDC) architectures. Recent work on QDCs, including Cisco-led studies, has investigated quantum data center networks with dynamically reconfigurable optical connections and benchmarked representative architectures such as QFly, BCube, Clos, and Fat-Tree under optical loss, contention, and scheduling constraints~\cite{Shapourian2025QDC,Pouryousef2026Benchmarking}. These architectures provide different tradeoffs between path length, path diversity, shared BSM resources, and switching overhead, all of which affect whether the black-box copy budgets computed in this paper can be supplied efficiently. In particular, BCube-like server-centric designs warrant further investigation because their structured path diversity can provide multiple end-to-end copies in proportion to the number of hops separating a QPU pair~\cite{Guo2009BCube}.

The present analysis focuses on resource allocation for a single demand, leaving concurrent-demand allocation as an important direction for future work. For a requested QPU pair, particularly one whose endpoints are separated by the network diameter, a topology-aware implementation may need to activate many candidate paths or reserve a substantial fraction of the available network resources to generate the number of raw end-to-end copies indicated by the black-box benchmark. In this sense, realizing the black-box copy budget may resemble a flooding or resource-reservation procedure for a single end-to-end demand. This abstraction is appropriate for quantifying the intrinsic resource requirements of hop-independent fidelity; however, quantum data centers must ultimately support many QPU--QPU demands concurrently. Multiple demands could be accommodated through time-division access, in which requests share the same physical resources across different time windows, or network-division access, in which disjoint network regions, path sets, wavelengths, or memory banks are assigned to different demands. Prior work on multi-flow entanglement routing and spatial and temporal multiplexing provides useful starting points for this extension~\cite{Pant2019Routing,Patil2021SpaceTimeMultiplexed}.

The purification strategy itself leaves room for improvement in both how purification is scheduled and when it is applied within the entanglement-distribution pipeline. Regarding the first question, this paper uses an all-in recursive schedule, in which all available copies are consumed as early as possible and successful outputs are pooled before the next purification level. This choice is motivated by the fact that the earliest levels act on the lowest-fidelity states and therefore experience the greatest attrition. However, the all-in schedule is not claimed to be globally optimal among all possible purification policies. More general adaptive schedules could determine whether to store, discard, regroup, or delay intermediate outputs based on the number of successes realized at each level. Such policies may reduce the required copy budget, particularly in regimes where success probabilities vary substantially across purification levels.

A complementary question is when, within the entanglement-distribution pipeline, purification should be applied. The present black-box model assumes that routing and entanglement swapping first generate raw end-to-end copies, which are then purified. This swap-then-purify abstraction is natural for isolating the end-to-end copy requirement. Alternatively, a network could purify elementary links or intermediate segments before swapping or adopt a hybrid strategy in which purification is applied at multiple stages. These approaches introduce different tradeoffs among memory time, classical-communication requirements, local-operation overhead, and the quality of the states entering subsequent swapping operations.

The present model treats all raw copies as identical Werner states generated over paths of the same length. This gives a clean benchmark, but real QDCNs will generally produce heterogeneous inputs: different paths may have different lengths, elementary links may have different fidelity, and storage time may vary across copies. A more complete model should therefore allow purification with unequal-quality inputs, imperfect Bell-state measurements, local gate noise, finite memory coherence, and nonuniform link-generation probabilities. These effects are likely to increase the required resources and may change which purification block sizes and recursion depths are optimal. Extending the dynamic-programming framework to heterogeneous input states would make the benchmark more directly applicable to device-level network simulations.

Finally, although the current analysis focuses on bipartite entanglement and discards elementary links that cannot contribute to end-to-end paths, a richer architecture could store and reuse leftover entanglement, or exploit multipartite states when available. Multipartite resources may be especially relevant in data-center settings, where many QPUs participate in distributed algorithms and the same physical fabric must support both pairwise and multipartite entanglement requests. Incorporating leftover bipartite and multipartite resources into future purification and routing decisions could further reduce the effective cost of hop-independent fidelity.

\section{Conclusion}
\label{sec:conclusion}

This paper developed a topology-independent end-to-end purification-resource benchmark
for hop-independent fidelity in multi-hop quantum data-center networks. We
modeled each elementary link as a Werner state with Werner parameter $w_0$ and
treated an $\ell$-link end-to-end connection as providing raw copies with
equal Werner parameter $w_0^\ell$. The goal was to determine the minimum number of
raw end-to-end copies needed to produce at least one purified output with
Werner parameter $w_{\mathrm{out}}\geq w_0$ and overall success probability at least
$p_{\mathrm{th}}$.

We compared recursive BBPSSW purification with higher-order
$r$-to-$1$ bilocal-Clifford purification protocols using an all-in recursive
schedule. The success probability of this schedule was computed by exact
dynamic programming over the number of surviving copies at each
purification level. The resulting resource landscapes show that
fulfilling hop-independent fidelity conditions has a threshold-like structure governed by
the Werner parameter of the raw end-to-end states, $w_0^\ell$, and the entanglement separability boundary
$w_0^\ell>1/3$. Increasing the success threshold mainly increases the raw
copy budget after fidelity recovery becomes possible, while the existence of
a feasible solution is controlled primarily by the raw end-to-end state's Werner parameter and
the purification dynamics.

Across the evaluated regimes, higher-order multi-copy purification
substantially reduces the required copy budget relative to recursive
BBPSSW and achieves fidelity recovery with shallower purification schedules. These
results identify the purification-side resource requirements that future
QDC architectures must meet through topology, multiplexing, repeated
connection attempts, memory-assisted scheduling, or combinations of these
mechanisms. They therefore provide a quantitative starting point for
designing network architectures that can support hop-independent
end-to-end entanglement distribution in practice.

\section*{Acknowledgment}

KPS thanks the U.S. Department of Energy, Office of Science, Advanced Scientific Computing Research (ASCR) program, for support under Award Number DE-SC0026264, and PQI Community Collaboration Awards. This research was also supported in part by the University of Pittsburgh Center for Research Computing and Data (RRID: SCR 022735) through the computational resources provided. The work utilized the HTC clusters, which are supported by NIH award number S10OD028483 and NSF award number OAC-2117681.

\bibliographystyle{IEEEtran}
\bibliography{qce_intro_core}

\end{document}